\documentstyle[12pt]{article}
              
\title{On the Quantum Space-Time Coordinates of an Event.}
\author{M. Toller   \\ 
Dipartimento di Fisica dell'Universit\`a, Trento  \\
I.N.F.N. gruppo collegato di Trento, Italia}
 
\newtheorem{proposition}{Proposition}
\newtheorem{assumption}{Assumption}
\begin{document} 
\maketitle                             
                 
\begin{abstract}
The present paper deals with the quantum coordinates of an event in spacetime, individuated by a quantum object. It is known that these observables cannot be described by selfadjoint operators. We describe them by means of a normalized positive operator valued (POV) measure in the Minkowski spacetime, satisfying a suitable covariance condition with respect to the Poincar\'e group.  This POV measure determines the probability that a measurement of the coordinates of the event gives results belonging to a given set in spacetime.  A general expression for these normalized covariant POV measures is given.

\end{abstract}

\newpage

\section{Introduction.}  

A quantum frame \cite{AK, Mayburov, Rovelli, Toller} is a material quantum object that individuates, within the accuracy permitted by the indeterminacy relations, a frame of reference in the Minkowski space-time. The observables which describe the relations between two quantum frames are the quantum analogs of the parameters which label an element of the Poincar\'e group. The algebra generated by them has been discussed in ref.\ \cite{Toller}. In the present paper we consider a limit case, namely the relations between a quantum frame and a classical frame. Moreover, we limit our attention to the four coordinates  $X^{\alpha}$, ${\alpha = 0, 1, 2, 3}$, which determine the origin of the quantum frame with respect to the classical frame. We may say that these four observables determine a ``quantum event''. It has been stressed in ref.\ \cite{Toller} that a quantum system must satisfy some conditions in order to define a quantum frame. In order to define a quantum event, it has to satisfy a weaker condition that we shall specify in the following. 

It is natural to assume that in a suitable dense domain of the Hilbert space $\cal H$
we have ($\hbar = 1$, $g^{00}=1$)
\begin{equation} \label{Commu}
[P^{\alpha}, X^{\beta}] = i g^{\alpha \beta}, 
\end{equation}  
or, in the domain where the operators $X^{\alpha}$ are defined, 
\begin{equation} \label{Commu2}
\exp(-ix_{\alpha} P^{\alpha}) X^{\beta} \exp(ix_{\alpha} P^{\alpha}) =  X^{\beta} +x^{\beta}. 
\end{equation}
If the operator $p_{\alpha} X^{\alpha}$ is self-adjoint, we have
\begin{equation}
\exp(ip_{\alpha} X^{\alpha}) P^{\beta} \exp(-ip_{\alpha} X^{\alpha}) =  P^{\beta} +p^{\beta} 
\end{equation} 
and it follows that the joint spectrum of the four-momentum operators $P^{\alpha}$ is invariant under translations in the direction of the four-vector $p^{\alpha}$. Since this joint spectrum is contained in the future cone, $p_{\alpha} X^{\alpha}$ cannot be self-adjoint. It follows that the operators  $X^{\alpha}$ cannot have a spectral representation and the statistical interpretation of the corresponding observables requires some particular attention.
                      
The argument given above, discussed by Wightman \cite{Wightman}, is an immediate generalization of a well known argument due to Pauli \cite{Pauli} concerning the time observable $T$, namely the quantity obtained by reading a quantum clock. It satisfies the commutation relation
\begin{equation} 
{ d T \over dt} =  i [H, T] = 1, 
\end{equation} 
where $t$ is the usual time parameter, measured by a classical external clock. If $T$ is self-adjoint this equation contradicts the fact that the spectrum of the Hamiltonian $H$ is bounded from below. 

Our coordinate $X^0$ is strictly related to the reading of a clock, but it is more similar to a time-of-arrival observable \cite{Allcock, BJ, GRT, Leon}, namely the time registered by a classical clock when some event happens, for instance a quantum particle reaches a given point, or two quantum particles collide. If we consider a quantum clock, the time-independent observable
\begin{equation}
X^0 = t - T
\end{equation}
is the time t measured by a classical clock when the quantum clock gives  $T = 0$, and it is a typical time-of-arrival observable.  Its commutator with the Hamiltonian $H = P^0$ is given by eq.\ (\ref{Commu}). Here we deal with an ``indirect'' measurement of a time-of-arrival, namely the measurement operation can be performed at any time $t$ and we use the equations of motion, which are supposed to be known. A different and more difficult problem is the ``direct'' measurement of a time-of-arrival, performed by means of operations lasting a long time and detecting immediately the event at the time at which it happens.
 
Several authors \cite{AB, Allcock, Armstrong, BJ, EF, GRT, JR, Leon, ORG, Pauli, Rosenbaum,  Rovelli2, SaW} have discussed the quantum time problem. A satisfactory solution has been given \cite{BGL, Giannitrapani, Holevo} by writing a generalized spectral representation 
\begin{equation} 
T = \int t \, d\tau(t), 
\end{equation}
where $\tau$ is a normalized positive operator valued (POV) measure on the real line. Since T is not selfadjoint, $\tau$ cannot be a projection valued measure (for a different point of view, see \cite{GRT}). The POV measure $\tau$ is not uniquely determined by the operator $T$, but it describes the time observable completely, since the probability that the result of a time measurement is contained in an interval I is given by
\begin{equation} 
{\cal P}(I) = (\psi , \tau(I) \, \psi), 
\end{equation} 
\begin{equation} 
\tau(I) =  \int_I d\tau(t), 
\end{equation} 
where the normalized vector $\psi$ describes the quantum state of the clock.

Observables of this kind have been considered for different purposes by several authors (see \cite{BLM, Davies, Holevo, Ludwig2}, where references to the original papers can be found).  The operator $\tau(I)$ represents a test \cite{Giles, Ludwig}, namely a (possibly mixed) yes-no observable. If we decompose the real line into a set of non-overlapping intervals $I_1,\ldots, I_n$, the operators $\tau(I_1),\ldots, \tau(I_n)$ represent a multi-bin test. One can show \cite{Lubkin, Lubkin2} that for any multi-bin test one can find a corresponding measuring instrument, if there are no limitations to the choice of the interaction Hamiltonian. This result legitimates the use of observables defined by POV measures within the standard formalism of quantum theory.

The aim of the present paper is to use a POV measure in order to describe all the space-time coordinates $X^{\alpha}$ of an event  measured with respect to a classical reference frame.  The quantities $X^1, X^2, X^3$ should not be confused with the self-adjoint Newton-Wigner coordinates of a particle \cite{NW, Wightman}, which do not commute with the Hamiltonian $P^0$, since the position of the particle changes with time. The coordinates of an event are clearly time-independent.

The final motivation of this research is to prepare a consistent formalism for a discussion of the limitations due to quantum gravity to the measurements of time and position \cite{DFR,  Ferretti, Garay, Mead}. One would like to show that the quantum coordinates of an event cannot be determined with a precision better than the Planck length, but a clear treatment is not possible in the absence of a clear definition of the quantum coordinates.

In Sections 2 and 3 we discuss the properties of the POV measures in the Minkowski space-time which are normalized and covariant with respect to the Poincar\'e group.  In Section 4 we give an explicit general formula for these POV measures. It is expected that, given a suitable physical object, the choice of the POV measure is not uniquely determined. In fact there is a large arbitrariness in the choice of the conventions which define the event in terms of the properties and the motion of the object. Some criteria which permit to choose POV measures with a more specific physical meaning will be discussed in a forthcoming paper. In Section 5 we discuss briefly the analogous treatment of covariant normalized POV measures in Minkowski spaces of general dimension $d$. For $d = 1$ we obtain a known general formula \cite{Holevo} for all the normalized POV measures on the time axis which are covariant under the time translations.  In Section 6 we suggest how to perform the extension of the formalism to non-normalized POV measures, describing situations in which it is not certain that the event really happens.

\bigskip   

\section{POV measures in space-time.}  

In order to describe the space-time coordinates of an event, we consider a normalized POV measure $\tau(I)$ on the Minkowski space-time $\cal M$. If the normalized vector $\psi$ in the separable Hilbert space $\cal H$ describes the state of the system that defines the event, the probability that the event is found in the Borel set $I \subset {\cal M}$ is given by
\begin{equation} 
{\cal P}(I) = 
(\psi, \tau(I) \psi). 
\end{equation} 
The normalization condition is
\begin{equation} \label{Normaliz}
\tau({\cal M}) = 1.
\end{equation} 
Then we put  
\begin{equation} \label{Coordinates}
X^{\alpha} = \int_{\cal M} x^{\alpha} \, d\tau(x). 
\end{equation}
Since these operators cannot be self-adjoint, $\tau$ cannot be a projection valued measure. 

We indicate by $\tilde{\cal P}$ the universal covering of the proper orthochronous Poincar\'e group $\cal P$. For its elements we use the notation $(x, a)$, where $x$ is a four-vector which describes a translation  and $a \in SL(2, C)$. $\Lambda(a)$ is the $4 \times 4$ Lorentz matrix corresponding to $a$. If $U(x, a)$ is the unitary representation of $\tilde{\cal P}$ that acts on the space $\cal H$, we require that 
\begin{equation}  \label{Covariance}
U^{\dagger}(x, a) \tau(\Lambda(a)I + x) U(x, a) = \tau(I).
\end{equation} 
This means that the POV measure $\tau$ and the representation $U$ of $\tilde{\cal P}$ form a ``system of covariance''. If $\tau$ were a projection valued measure, we should have a ``system of imprimitivity'' \cite{Mackey}. If we consider translations, eq.\ (\ref{Covariance}) takes the form
\begin{equation} \label{Covariance2}
\exp(-ix_{\alpha} P^{\alpha}) \tau(I + x) \exp(ix_{\alpha} P^{\alpha}) = \tau(I).
\end{equation} 
The formula (\ref{Commu2}) follows from this equation and eq.\ (\ref{Coordinates}). Of course, the covariance assumption is valid if no external objects intervene in the definition of the event.

A consequence of eq.\ (\ref{Covariance2}) and of the spectrum of four-momentum is the following:  \begin{proposition} \label{Positivity}
If $\psi \neq 0$ and $I \subset {\cal M}$ is a non-empty open set, we have 
\begin{equation} 
(\psi, \tau(I) \psi) > 0.
\end{equation}
\end{proposition}   

Assume that this expression vanishes and find two open sets  $I'$ and $I''$  with the property  
\begin{equation} 
I' + I'' \subset I.
\end{equation}
Then we have
\begin{displaymath}
(\psi, \tau(I' + x) \psi) =
(\exp(-ix_{\alpha} P^{\alpha}) \psi,  \tau(I') \exp(-ix_{\alpha} P^{\alpha}) \psi) = 0 
\end{displaymath}
\begin{equation} 
{\rm for} \quad x \in I'',
\end{equation}   
namely
\begin{equation} 
(\tau(I'))^{1/2} \exp(-ix_{\alpha} P^{\alpha}) \psi = 0 
\qquad {\rm for} \quad x \in I''.
\end{equation}  
This expression is the limit of a vector-valued function analytic in the tube defined by ${\rm Im}\, x \in -V_0$, where $V_0$ is the open future cone. An application of the edge-of-the-wedge theorem \cite{StW} shows that this analytic function vanishes in the whole tube and it follows that 
\begin{equation} 
(\psi, \tau(I' + x) \psi) = 0 
\end{equation}
for any real value of $x$. Then we have $\tau = 0$ in contradiction with our assumption that the measure $\tau$ is normalized. 

It follows from Proposition \ref{Positivity} that, if $I$ has a non-empty interior, $\tau(I)$ cannot be a projection operator different from $1$.  We obtain in this way another proof that $\tau$ cannot be a projection valued measure. We also see that the localization of an event in a bounded region $I$ of space-time cannot be considered as a ``property'' of the state $\psi$.

It is clear that the covariance and the normalization conditions do not determine the POV measure $\tau$ uniquely. For instance, if $K$ is an unitary operator that commutes with all the operators $U(x, a)$, the POV measure
\begin{equation} 
\tau'(I) = K^{\dagger}\tau(I)K
\end{equation}
satisfies the required conditions as well as $\tau$. 

A system which defines a quantum event has to satisfy some conditions. We decompose the unitary representation $U(x, a)$ into a direct integral of irreducible unitary representations (IURs) of $\tilde {\cal P}$ \cite{Wigner}. Of course, only positive-energy representations appear in this decomposition. Moreover, we require that the following assumption is satisfied:
\begin{assumption} \label{Spectrum}
In the decomposition of $U(x, a)$ only a continuous mass spectrum appears, namely $U(x, a)$ has no irreducible subrepresentation.
\end{assumption}
As a consequence of this assumption, we need to consider only positive-mass IURs. 

If the theory we are considering does not contain zero mass particles and is asymptotically complete \cite{Jost, StW}, we can describe its states in terms of asymptotic ``in'' or ``out'' states and the Hilbert space of the theory can be written, for instance, as
\begin{equation}
\hat{\cal H} = {\cal H}_{\rm out} = {\cal H}^{(0)} \oplus  {\cal H}^{(1)} \oplus {\cal H}^{(2)}_{\rm out} \oplus \ldots,
\end{equation} 
where ${\cal H}^{(0)}$ contains the vacuum state, ${\cal H}^{(1)}$ contains states with one particle (or one stable bound state), ${\cal H}^{(2)}_{\rm out}$ contains states with two outgoing particles, and so on. The Assumption \ref{Spectrum} means that we have to consider states orthogonal to the vacuum and to the one-particle states. They form a subspace ${\cal H} \subset \hat{\cal H}$.  It is physically intuitive that the vacuum or a single particle cannot be used to define an event: at least two particles are necessary for this purpose. 

The description of $\cal H$ in terms of asymptotic states is physically interesting, because it deals with a situation in which the spacetime position of an event, for instance a collision, is measured by means of operations performed in a far-away region, as it happens, for instance, in the famous conceptual position measurement by means of a microscope, discussed by Heisenberg
\cite{Heisenberg} and reconsidered by Mead \cite{Mead} in the presence of the gravitational interaction. This point of view could also provide the starting point for the introduction of space-time concepts in a pure S-matrix theory.

We stress that we are dealing with ``indirect'' measurements of the coordinates of an event, namely the test $\tau(I)$ is not measured by means of physical operations performed in the space-time region $I$. For this reason we do not require that the operators $\tau(I)$ and $\tau(I')$ commute if the regions $I$ and $I'$ are space-like separated.

It is convenient to describe the IURs of $\tilde {\cal P}$ defined by Wigner \cite{Wigner}, corresponding to a mass $\mu > 0$ and ``spin'' (namely angular momentum in the centre of mass) $j$, by means of the formalism of induced representations \cite{Mackey}. The representation operators act on wave functions with $(2j + 1)$ components defined on $SL(2, C)$ and satisfying the covariance condition
\begin{equation} \label{Covar}
\psi_m(au) = \sum_{m'} R^j_{mm'}(u^{-1}) \psi_{m'}(a), \qquad u \in SU(2),
\end{equation} 
where $R^j$ is the well known $(2 j + 1)$-dimensional IUR of $SU(2)$.
If we choose for each four-momentum $k$ on the mass shell
\begin{equation}
k_{\alpha} k^{\alpha} =\mu^2, \qquad k^0 > 0,
\end{equation} 
an element $a_k \in SL(2, C)$ with the property 
\begin{equation}
k = \Lambda(a_k)q(\mu), \qquad  q(\mu) = (\mu, 0, 0, 0),
\end{equation}
the Wigner wave function is 
\begin{equation}
\psi_m(k) = \psi_m(a_k).
\end{equation}
and the norm is given by
\begin{equation} \label{Norm}
\|\psi\|^2 = \int \sum_m |\psi_m(a_k)|^2 \delta(k_{\alpha} k^{\alpha} - \mu^2) \theta(k^0) \, d^4 k.  
\end{equation}
Note that, as a consequence of the covariance condition (\ref{Covar}), the integrand function in eq.\ (\ref{Norm}) does not depend on the choice of the element $a_k$.  The representation operator is given by  
\begin{equation} \label{Repres}
[U(x, b)\psi]_m(a) = \exp(ik_{\alpha} x^{\alpha}) \psi_m(b^{-1}a), \qquad  
k = \Lambda(a) q(\mu).
\end{equation}

The Hilbert space ${\cal H}$ can be decomposed (in a non unique way) into a direct integral of spaces in which IURs of $\tilde{\cal P}$ operate. A vector $\psi \in {\cal H}$ is described by a wave function of the kind  $\psi_{\sigma j m}(\mu, a)$, where the index $\sigma$ labels the spaces in which equivalent IURs operate. It satisfies a covariance condition similar to eq.\ (\ref{Covar}) and it vanishes if $\mu$ does not belong to a $\sigma$-dependent mass spectrum. The norm is given by 
\begin{equation} \label{Norm2}
\|\psi\|^2 = 
\int_V \sum_{\sigma j m} |\psi_{\sigma j m}(\mu, a_k)|^2  \, d^4 k,  
\end{equation}
where $V$ is the future cone. Note that the integration over the four-vector $k$ implies an integration over the mass $\mu$. The representation operator is defined by a direct generalization of eq.\ (\ref{Repres}).  

\bigskip      

\section{Construction of covariant POV measures.}  

In this and in the following Section, we derive a representation for an arbitrary normalized covariant POV measure on the Minkowski space-time. The general properties of covariant POV measures on various kinds of spaces have been studied by many authors \cite{CH, Holevo, Holevo2, Kholevo, Scutaru} and their results could have been used in the first steps of our treatment. We prefer to give a self-contained, simple and constructive approach to our particular problem. We consider the dense Poincar\'e invariant linear space ${\cal D} \subset {\cal H}$ composed of the wave functions in four-momentum space which are $C^{\infty}$, have compact support and do not vanish only for a finite number of choices of the indices $\sigma, j, m$.  They have the property
\begin{equation} \label{SemiNorms}
\|\psi\|_r^2 = 
\int_V \sum_{\sigma j m} |(1+k_0)^r \psi_{\sigma j m}(\mu, a_k)|^2  \, d^4 k < \infty,
\qquad r = 0, 1,\ldots, 
\end{equation}
which implies that $F(P) \psi \in {\cal H}$ for any choice of the polynomial function  $F(P)$ of the four-momentum operators. The topology of $\cal D$ is the usual one \cite{Schwartz} and eq.\  (\ref{SemiNorms}) defines a  family of  continuous norms.

 The convolution of the numerical measure $(\psi, \tau(x) \phi)$ with the function $g(x)$, continuous and with compact support, is a function of $x$ given by 
\begin{displaymath}
\left(\psi, \int g(x - y) \, d \tau(y) \, \phi \right) = 
\end{displaymath}
\begin{equation} 
= \left(\exp(-ix_{\alpha} P^{\alpha}) \psi, \int g(-y) \, d \tau(y) \, \exp(-ix_{\alpha} P^{\alpha}) \phi \right).
\end{equation}
Its partial derivatives are given by sums of similar expressions in which the vectors $\psi$ and $\phi$ are replaced by vectors of the kind  $F(P) \psi$ and $F'(P) \phi$, where $F(P)$ and $F'(P)$ are polynomials. If $\psi, \phi \in {\cal D}$, we see that the convolution defined above is infinitely differentiable for any choice of the continuous function $g$.  A general theorem concerning distributions \cite{Schwartz} permits one to draw the following conclusion:
\begin{proposition} 
If $\psi, \phi \in {\cal D}$, we can write  
\begin{equation}  \label{Integral}
(\psi, \tau(I) \phi) = \int_I \rho(\psi, \phi, x) \, d^4 x, 
\end{equation} 
where $\rho(\psi, \phi, x)$ in an infinitely differentiable function of $x$. In particular
\begin{equation} \label{Integral2} 
(\psi, \tau(I) \psi) = \int_I \rho(\psi, x) \, d^4 x, 
\end{equation}
where
\begin{equation}  
\rho(\psi, x) = \rho(\psi, \psi, x) \geq 0.
\end{equation}
\end{proposition}  

If we introduce the set
\begin{equation}  
I(x) =\{y \in {\cal M}:  y^0 < x^0, y^1 < x^1, y^2 < x^2, y^3 < x^3, \}
\end{equation} 
we have
\begin{displaymath}
 \rho(\psi, \phi, x) = {\partial^4 \over \partial x^0 \partial x^1 \partial x^2 \partial x^3}  (\psi, \tau(I(x)) \phi)  =
\end{displaymath}
\begin{equation}  
= {\partial^4 \over \partial x^0 \partial x^1 \partial x^2 \partial x^3} \left(\exp(-ix_{\alpha} P^{\alpha}) \psi, \tau(I(0)) \exp(-ix_{\alpha} P^{\alpha}) \phi \right).
\end{equation} 
A simple calculation gives
\begin{equation}  
| \rho(\psi, \phi, x) | \leq  16 \|\psi\|_4 \|\phi\|_4
\end{equation}   
and we see that $\rho(\psi, \phi, x)$ for fixed values of $x$ is a continuous function of $\psi, \phi \in {\cal D}$.

The normalization and the covariance conditions take the form
\begin{equation} \label{Normalization}
\int \rho(\psi, x) \, d^4 x =  \| \psi \|^2, 
\end{equation} 
\begin{equation}  \label{Covariance3}
\rho(U(y, a) \psi, U(y, a) \phi, \Lambda(a) x + y) = \rho(\psi, \phi, x).
\end{equation} 
Since $\rho(\psi, \phi, x)$ is a continuous function of $x$, it has a well defined value at $x = 0$, which satisfies the invariance property
\begin{equation} \label{Invariance}
\rho(U(0, a) \psi, U(0, a) \phi, 0) = \rho(\psi, \phi, 0).
\end{equation}
If $\rho(\psi, \phi, 0)$ is given, we put 
\begin{equation} \label{Rhox}
\rho(\psi, \phi, x) = \rho(U(-x, 1) \psi, U(-x, 1) \phi, 0)
\end{equation}                       
and the covariance condition (\ref{Covariance3}) is satisfied.

Note that $\rho(\psi, \phi, 0)$ is a continuous positive sesquilinear form on the space $\cal D$, invariant under $SL(2, C)$.  It defines a scalar product in the quotient space ${\cal D}/{\cal D}_0$ and on its completion $\tilde{\cal H}$, which is a Hilbert space.  We have indicated by  ${\cal D}_0$ the subspace of ${\cal D}$ defined by the condition  $\rho(\psi, 0) = 0$. It is invariant under $SL(2, C)$. This construction also defines a linear operator $h: {\cal D} \to \tilde{\cal H}$ and we have
\begin{equation} \label{ScalProd}
\rho(\psi, \phi, 0) = (h\psi, h\phi),
\end{equation} 
where at the right hand side there is the scalar product of the Hilbert space $\tilde{\cal H}$. $SL(2,C)$ acts on $\tilde{\cal H}$ by means of the unitary representation $a \to \tilde U(a)$, which has the property
\begin{equation} 
\tilde U(a) h = h U(0, a).
\end{equation}
This means that $h$ is an intertwining operator. It is continuous, since we have 
\begin{equation}  \label{Cont}
\| h \psi \| \leq 4 \| \psi\ \|_4.
\end{equation}

The unitary representation $\tilde U$ and the intertwining operator $h$ determine the POV measure $\tau$ by means of eqs.\ (\ref{Integral}), (\ref{Rhox}) and (\ref{ScalProd}). In order to describe the representation $\tilde U$, we shall consider its direct integral decomposition into IURs of $SL(2, C)$ and the corresponding decomposition of the Hilbert space $\tilde{\cal H}$  into a direct integral of irreducibles spaces labelled by the variable $\gamma \in \Gamma$
\begin{equation} \label{DirInt}
\tilde{\cal H} = \int^{\oplus}_{\Gamma} \tilde{\cal H}_{\gamma} \, d\omega(\gamma).
\end{equation}                    
The variable $\gamma$ contains the parameters that label the equivalence classes of IURs of $SL(2, C)$ and an index $\nu$ that distinguishes the spaces where equivalent IURs operate. The measure $\omega$ is positive and we always disregard subsets of $\Gamma$ with vanishing measure. In the simplest and physically most interesting cases, the measure $\omega$ is discrete and $\tilde{\cal H}$ is decomposed into a direct sum of irreducible subspaces. 

 In the next Section, taking also into account the normalization condition, we find the general form of the intertwining operator $h$ when the representation $\tilde U$ is given.

\bigskip

\section{Use of the IURs of $SL(2,C)$.}  
 
In order to obtain more explicit formulas, we have to use the detailed properties of the IURs of $SL(2,C)$. Their matrix elements $D^{M c}_{jmj'm'}(a)$  are treated in refs.\ \cite{GGV, Naimark, Ruhl}. We adopt the conventions of ref.\ \cite{ST}, where explicit formulas for these quantities can be found (the parameter $c$ is called $\lambda$ in ref.\ \cite{ST}). There are two series of IURs: the principal series with $c$ imaginary and $M$ integral or half-integral, and the supplementary series with $-1 < c <1$ and $M = 0$. The representations  $D^{M c}$ and $D^{-M -c}$ are unitarily equivalent. The restriction of these representations to the subgroup $SU(2)$ is given by
\begin{equation} 
D^{M c}_{jmj'm'}(u) = \delta_{jj'} R^j_{mm'}(u), \qquad u \in SU(2)
\end{equation}    
and the possible values of $j$ are
\begin{equation}  
j = |M|, |M| + 1,\ldots.
\end{equation}
For the invariant measures on $SL(2, C)$ and $SU(2)$ we use the normalization
\begin{equation}  
\int_{SL(2, C)} f(a) \, d^6 a =  \mu^{-2} \int_{SU(2)} d^3 u \int f(a_k u) \, \delta(k_{\alpha} k^{\alpha} - \mu^2) \theta(k^0) \, d^4 k, 
\end{equation}
\begin{equation} 
\int_{SU(2)} d^3 u = 1.
\end{equation} 

The index $\gamma \in \Gamma$, which appears in the direct integral decomposition (\ref{DirInt}), stands for the discrete  indices $\nu, M$ and for the continuous parameter $c$. An element $\Psi \in \tilde{\cal H}$ can be described by the quantity $\Psi_{\gamma ln}$. Its norm is given by
\begin{equation} 
\|\Psi\|^2 = \int_{\Gamma} \sum_{ln} |\Psi_{\gamma ln}|^2 \, d \omega(\gamma)
\end{equation}
and the representation $\tilde U$ acts in the following way
\begin{equation} 
[\tilde U(a) \Psi]_{\gamma ln} = \sum_{l'n'} D^{M c}_{lnl'n'}(a) \Psi_{\gamma l'n'},
\end{equation} 
where the parameters $M, c$ depend on $\gamma$.
 
From eqs.\ (\ref{Rhox}) and (\ref{ScalProd}) we have 
\begin{equation} \label{Density}
\rho(\psi, x) = \int_{\Gamma} \sum_{ln} |\Psi_{\gamma ln}(x)|^2 \, d \omega(\gamma),
\end{equation} 
where 
\begin{equation} \label{PsiX}
\Psi(x) = h \exp(-i x^{\alpha} P_{\alpha}) \psi. 
\end{equation}               
The normalization condition takes the form
\begin{equation} \label{Normal}
\int \int_{\Gamma} \sum_{ln} |\Psi_{\gamma ln}(x)|^2 \, d \omega(\gamma) \, d^4 x =  \|\psi\|^2
\end{equation} 
and we see that eq.\ (\ref{PsiX}) defines an isometric linear mapping $\hat h: {\cal D} \to \tilde{\cal H} \otimes L_2({\cal M})$, which can be extended by continuity to the whole space $\cal H$. It follows that the probability density $\rho(\psi, x)$ is an integrable function (in general not continuous) for all the vectors $\psi \in {\cal H}$. 

We also consider the isometric linear mapping $\tilde h$ that transforms the vector $\psi$ into the Fourier transform
\begin{equation} 
\tilde\Psi_{\gamma ln}(k) = (2 \pi)^{-2} \int \exp(i x_{\alpha} k^{\alpha}) \Psi_{\gamma ln}(x)\, d^4 x.
\end{equation}
We can easily see that
\begin{equation} 
[\tilde h \exp(-i y_{\alpha} P^{\alpha}) \psi]_{\gamma ln}(k) =
\exp(-i y_{\alpha} k^{\alpha}) [\tilde h \psi]_{\gamma ln}(k), 
\end{equation}
namely that the operator $\tilde h$  commutes with the translations. It follows that it is decomposable \cite{Dixmier}, namely it has the following form diagonal with respect to the variable $k$ 
\begin{equation} 
\tilde \Psi_{\gamma ln}(k) = \sum_{\sigma j m} K_{\gamma ln \sigma j m}(k) \psi_{\sigma j m}(\mu, a_k).
\end{equation}
By inverting the Fourier transformation, we obtain
\begin{equation} \label{PsiX2}
\Psi_{\gamma ln}(x) = (2 \pi)^{-2} \int_V \exp(-i x_{\alpha} k^{\alpha}) \sum_{\sigma j m} K_{\gamma ln \sigma j m}(k) \psi_{\sigma j m}(\mu, a_k) \, d^4 k.
\end{equation} 

In particular, we have the following representation of the operator $h$:
\begin{equation} 
[h \psi]_{\gamma ln} = \Psi_{\gamma ln}(0) = (2 \pi)^{-2} \int_V \sum_{\sigma j m} K_{\gamma ln \sigma j m}(k) \psi_{\sigma j m}(\mu, a_k) \, d^4 k.
\end{equation} 
If we put  
\begin{equation}  \label{Kernel}
K_{\gamma ln \sigma j m}(\mu, a_k u)  =  \sum_{m'}  K_{\gamma ln \sigma j m'}(k) R^j_{m' m}(u),
\end{equation}
we can also write
\begin{equation} 
\Psi_{\gamma ln}(0) = (2 \pi)^{-2} \int \sum_{\sigma j m} K_{\gamma ln \sigma j m}(\mu, a) \psi_{\sigma j m}(\mu, a) 2\mu^3 \, d \mu \, d^6 a.
\end{equation} 

From the intertwining property, using the invariance of the measure $d^6 a$, we obtain   
\begin{equation} 
\sum_{l' n'} D^{M c}_{lnl'n'}(b)  K_{\gamma l'n' \sigma j m}(\mu, a) = K_{\gamma ln \sigma j m}(\mu, ba).
\end{equation}
 It follows that
\begin{equation} 
\sum_{l' n'} D^{M c}_{lnl'n'}(a^{-1})  K_{\gamma l'n' \sigma j m}(\mu, a) = \delta_{lj} \delta_{nm} F^j_{\gamma \sigma}(\mu). 
\end{equation}
In fact, the left hand side does not depend on $a$ and the structure of the right hand side follows from the invariance under the substitution $a \to au$ with $u \in SU(2)$ and from eqs.\ (\ref{Covar}) and (\ref{Kernel}). Then we can write
\begin{equation} 
K_{\gamma ln \sigma j m}(\mu, a) = D^{M c}_{lnjm}(a) F^j_{\gamma \sigma}(\mu), 
\end{equation} 
where $F^j_{\gamma \sigma}(\mu)$ vanishes unless $j = |M(\gamma)|, |M(\gamma)| + 1,\ldots$.

From these results, we obtain
\begin{equation}  \label{Wave}
\Psi_{\gamma ln}(x) = (2 \pi)^{-2}  \int_V \exp(-i x_{\alpha} k^{\alpha}) \sum_{\sigma j m} D^{M c}_{lnjm}(a_k) F^j_{\gamma \sigma}(\mu) \psi_{\sigma j m}(\mu, a_k) \, d^4 k
\end{equation}
and after some calculation we have
\begin{equation}
\int \rho(\psi, x) \, d^4 x = \int_{\Gamma} \int_V \sum_{jm} \left|\sum_{\sigma} F^j_{\gamma \sigma}(\mu) \psi_{\sigma j m}(\mu, a_k) \right|^2 \, d^4 k \, d\omega(\gamma).
\end{equation}
By comparison with eq.\ (\ref{Norm2}) we see that the normalization condition (\ref{Normalization}), is satisfied if
\begin{equation} \label{Condition}
 \int_{\Gamma} \overline{F_{\gamma \sigma}^j(\mu)} F_{\gamma \sigma'}^j(\mu) \, d\omega(\gamma) = \delta_{\sigma \sigma'}.
\end{equation} 

In conclusion, we have seen that
\begin{proposition}
The the most general normalized covariant POV measure $\tau$ on the Minkowski space is given by eq.\ {\rm (\ref{Integral2})} in terms of the density $\rho(\psi, x)$ which is an integrable function for any $\psi \in {\cal H}$. For $\psi \in {\cal D}$ this density is given by eqs.\ {\rm (\ref{Density})} and {\rm (\ref{Wave})} in terms of the space $\Gamma$, the measure $\omega$, and the function $F_{\gamma \sigma}^j(\mu)$ satisfying the condition {\rm (\ref{Condition})}. 
\end{proposition} 
 
If the measure $\omega$ is discrete and $\omega(\{\gamma\}) = 1$, we have
\begin{equation}
\rho(\psi, x) = \sum_{\gamma ln} |\Psi_{\gamma ln}(x)|^2 
\end{equation} 
and the  condition (\ref{Condition}) can be written in the simpler form
\begin{equation} 
 \sum_{\gamma} \overline{F_{\gamma \sigma}^j(\mu)} F_{\gamma \sigma'}^j(\mu) = \delta_{\sigma \sigma'}.
\end{equation} 
It means that, for fixed values of $j$ and $\mu$, the matrix $F_{\gamma \sigma}^j(\mu)$ is isometric.  

It is interesting to remark that the IURs of $SL(2, C)$ belonging to the supplementary series may appear in the decomposition of the unitary representation $\tilde U$ on which the construction of the POV measure is based. On the contrary, they do not appear in the direct integral decomposition, based on the Plancherel formula \cite{GGV, Naimark, Ruhl},  of the unitary representation $U(0, a)$ which acts on the physical Hilbert space $\cal H$.  This could not happen if the intertwining operator $h$, which is defined on $\cal D$, had an isometric extension to the whole Hilbert space $\cal H$. The existence of this extension, however, does not follow from our assumptions.
 
\bigskip 

\section{Lower-dimensonal space-times.}

The treatment given above can be generalized, with some evident modifications, to space-times with arbitrary dimension.  If we consider only the time dimension, the homogeneous Lorentz group and its unitary representation $\tilde U$ are trivial and the index $\gamma$ simply labels a basis in the Hilbert space $\tilde{\cal H}$. We describe the states, which must have a continuous energy spectrum, by means of wave functions of the kind $\psi_{\sigma}(E)$ with the norm
\begin{equation}
\|\psi\|^2 = \int \sum_{\sigma} |\psi_{\sigma}(E)|^2 \, dE.
\end{equation} 
Then our results take the following simple form, which is equivalent to a result given in ref.\ \cite{Holevo}: 
\begin{proposition} 
The most general normalized POV measure on the time axis covariant with respect to the time translations is given by the density
\begin{equation}
\rho(\psi, t) =  \sum_{\gamma} |\Psi_{\gamma}(t)|^2,
\end{equation}
where 
\begin{equation}  
\Psi_{\gamma}(t) = (2 \pi)^{-1/2}  \int \exp(-i t E) \sum_{\sigma}  F_{\gamma \sigma}(E) \psi_{\sigma}(E) \, d E
\end{equation}
and the function $F_{\gamma \sigma}(E)$ satisfies the condition 
\begin{equation} 
\sum_{\gamma} \overline{F_{\gamma \sigma}(E)} F_{\gamma \sigma'}(E) = \delta_{\sigma \sigma'}.
\end{equation}   
\end{proposition} 

Relevant simplifications also appear in $1+1$ dimensions, because the homogeneous Lorentz group is a one-parameter group, labelled by the rapidity parameter $\zeta$. Its IURs are one-dimensional and have the exponential form $\exp(i c \zeta)$. The wave functions have the form  $\psi_{\sigma}(\mu, \zeta)$, where
\begin{equation}
k^0 = \mu \cosh\zeta, \qquad  k^1 = \mu \sinh\zeta
\end{equation}
and the norm is given by 
\begin{equation}
\|\psi\|^2 = \int \sum_{\sigma} |\psi_{\sigma}(\mu, \zeta)|^2 \, \mu d\mu \, d\zeta.
\end{equation}
Then the formulas of the preceding Section take the simpler form
\begin{equation}
\rho(\psi, x) = \int_{\Gamma} |\Psi_{\gamma}(x)|^2 \, d \omega(\gamma),
\end{equation} 
\begin{displaymath}
\Psi_{\gamma}(x) =
\end{displaymath}
\begin{equation}  
 = (2 \pi)^{-1}  \int \exp(-i x_{\alpha} k^{\alpha})  \exp(i c \zeta) \sum_{\sigma} F_{\gamma \sigma}(\mu) \psi_{\sigma}(\mu, \zeta) \, \mu d\mu \, d\zeta,
\end{equation}
where $c$ depends on $\gamma$ and the function $F_{\gamma \sigma}(\mu)$ satisfies the condition 
\begin{equation} 
\int_{\Gamma} \overline{F_{\gamma \sigma}(\mu)} F_{\gamma \sigma'}(\mu) \, d\omega(\gamma) = \delta_{\sigma \sigma'}.
\end{equation}    

\bigskip 

 \section{Non-normalized POV measures.}
  
Sometimes it is necessary to consider the possibility that an event does not take place \cite{GRT}. This situation can be described by a non-normalized POV measure, namely the normalization condition (\ref{Normaliz}) has to be replaced by
\begin{equation} 
\|\tau({\cal M})\| \leq 1.
\end{equation}
The positive operator $\tau({\cal M})$, which commutes with all the representation operators $U(x, a)$, represents a test and the ``yes'' outcome means that the the event has taken place.  If it is a projection operator, we have just to replace the Hilbert space $\cal H$ by its subspace  $\tau({\cal M}) {\cal H}$. We have already done an operation of this kind when, in agreement with the Assumption \ref{Spectrum}, we have required that the vector $\psi$ is orthogonal to the vacuum and to the one-particle states. 

Even if $\tau({\cal M})$ is not a projection operator, the results obtained in the preceding Sections can easily be adapted to this more general situation. The operators $\hat h$ and $\tilde h$ introduced in Section 4  are bounded, but not necessarily isometric and the condition (\ref{Condition}) has to be replaced by   
\begin{equation} 
\sum_{\sigma \sigma'} \overline c_{\sigma} c_{\sigma'} \int_{\Gamma} \overline{F_{\gamma \sigma}^j(\mu)} F_{\gamma \sigma'}^j(\mu) \, d\omega(\gamma) \leq \sum_{\sigma} |c_{\sigma}|^2,
\end{equation} 
for any choice of the complex numbers $c_{\sigma}$.      

\bigskip  \bigskip

\noindent {\it Acknowledgments:} I am grateful to Dr.\ R. Giannitrapani for many useful discussions.

\end{document}